\documentclass{PoS}

\usepackage{graphicx}
\usepackage{epstopdf}

\title{Meson spectrum and tetraquarks\\through an AdS/QCD inspired potential}

\ShortTitle{Quark model and AdS/QCD}

\author{{Floriana Giannuzzi}\\
         Dipartimento di Fisica, Universit\`a degli Studi di Bari, I-70126 Bari, Italia\\
        I.N.F.N., Sezione di Bari, I-70126 Bari, Italia\\
        E-mail: \email{floriana.giannuzzi@ba.infn.it}}

\abstract{AdS/QCD correspondence allows to compute the quark-antiquark potential in the static
limit. We use this 
piece of information together with the Salpeter equation 
(Schroedinger equation with relativistic kinematics) and a short range hyperfine
splitting potential to describe the bound states of quark-antiquark.
Comparison with experiment permits to fit the parameters of the model, namely the quark masses and the quark potential parameters. A discussion of heavy tetraquark masses, 
prompted by the hypothesis of a diquark-antidiquark structure for the puzzling state X(3872), is also presented. }

\FullConference{8th Conference Quark Confinement and the Hadron Spectrum \\
		 September 1-6 2008\\
		 Mainz, Germany}

\begin{document}

%\section{...}
A new promising tool  to investigate the strong coupling regime of QCD is the AdS/QCD correspondence. In this framework many results have been obtained, namely linear  Regge trajectories, the glueball spectrum, the light meson spectrum and decay constants \cite{Karch:2006pv}.
Through the AdS/QCD correspondence it is also possible to determine the $Q\bar{Q}$ potential \cite{Andreev:2006ct}, the static energy of the strongly interacting quark-antiquark pair. 

This result can be used in a quark model  to get the meson spectrum \cite{Carlucci:2007um}; the application can be extended to compute some tetraquark spectra, under the hypothesis that a tetraquark can be considered as a bound state of a diquark and an antidiquark \cite{Maiani:2004vq}.

To compute the meson masses, we solve the semirelativistic wave equation (Salpeter equation): 
\begin{equation}\label{salpeq}
\left(\sqrt{m_1^2-{\bf \nabla}^2}+\sqrt{m_2^2-{\bf \nabla}^2}+V(r)\right) \psi({\bf r})=M\, \psi({\bf r})
\end{equation}
where $M$ and $\psi$ are the mass and the wavefunction of the meson, respectively, and $m_1$ and $m_2$ are the masses of the constituent quarks. Only the case $\ell=0$ has been considered. The potential $V(r)$ in (\ref{salpeq}) comprises three terms: 
\begin{equation}\label{fullpot}
V(r)=V_{AdS/QCD}(r)+V_{spin}(r)+V_0\,;
\end{equation}
the last term $V_0$ is a constant.
$V_{AdS/QCD}$ is the AdS/QCD inspired potential, obtained in \cite{Andreev:2006ct} as a parametric equation:
\begin{equation}\label{potadsqcd}\left\{
\begin{array}{cc}\label{zakpot} 
V_{AdS/QCD}(\lambda)\,=\,\frac{g}{\pi}
\sqrt{\frac{c}{\lambda}} \left( -1+\int_0^1{dv \, v^{-2} \left[
\mbox{e}^{\lambda v^2/2} \left(1-v^4
\mbox{e}^{\lambda(1-v^2)}\right)^{-1/2}-1\right]} \right) & \\
r(\lambda)\,=\,2\, \sqrt{\frac{\lambda}{c}} \int_0^1{dv\, v^{2}
\mbox{e}^{\lambda (1-v^2)/2} \left(1-v^4
\mbox{e}^{\lambda(1-v^2)}\right)^{-1/2}}  \hspace{3.3cm}&  \,,
\end{array}
\right.\end{equation}
where $r$ is the interquark distance and $\lambda$ varies in the range: $0\leq\lambda<2$. The potential $V_{AdS/QCD}(r)$ depends only on
two parameters, $g$ and $c$. Since the calculation in \cite{Andreev:2006ct} is performed in the static limit, which corresponds to large quark masses, only heavy meson masses, i.e. of mesons comprising at least one heavy quark, are computed. 
The second term in (\ref{fullpot}) takes account of the spin interaction and, in the  one-gluon-exchange approximation and in the case $\ell=0$, can be written as:
\begin{equation}
V_{spin}(r)=A \frac{\delta(r)}{m_1 m_2}({\bf S}_1\cdot {\bf S}_2) \qquad \mbox{with} \qquad \delta(r)=\left(\frac{\sigma}{\sqrt{\pi}}\right)^3 \mbox{e}^{-\sigma^2 r^2}. \nonumber
\end{equation}
$A$ is a constant  proportional to the QCD running coupling constant; to account for the different value of $\alpha_s$ at the two scales,  we have introduced two different parameters, $A_c$ in the case of mesons comprising at least one charm quark  and $A_b$ for bottom quark. 

We have also introduced a cutoff to cure the unphysical divergence at $r=0$ of the wavefunction solution of the Salpeter equation: at distances shorter than $r_M=k/M$, where $k$ is a constant and $M$ is the mass of the meson, the potential is taken constant, fixed at the value $V(r_M)$. The constant $k$ has different values in the case of mesons made up of identical quarks $(k)$ and mesons made up of different quarks $(k')$.

To solve the Salpeter equation, we use the Multhopp method, a numerical method  which allows to transform an integral equation into a set of linear equations, introducing $N$ parameters $\theta_k$, called Multhopp's
angles. The set of equations is (we refer to
\cite{Colangelo:1990rv} for further details):
\begin{equation}
\sum_{m=1}^{N}B_{km}\psi(\theta_k)\,=\,M\psi(\theta_m)
\end{equation} 
where $\psi(\theta_k)= u_0(-\cot\theta_k)$ and 
$\theta_k=\frac{k\pi}{N+1} \, \, (k=1,\cdots N)$.

Finally, we fix the parameters of the potential and the constituent quark masses using, as an input, some heavy meson masses. We obtain the following parameters: $c$=0.3  GeV$^2$, $g$=2.75, $V_0$=-0.49  GeV,
$k$=1.48, $k'$=2.15, $A_c$=7.92, $A_b$=3.09, $\sigma$=1.21  GeV, $m_q$=0.302
 GeV, $m_s$=0.454  GeV , $m_c$=1.733 
GeV, $m_b$=5.139  GeV. Using these values,  we find the meson spectrum reported in Table \ref{tmesons}.

\begin{table}[ht!]
\caption{\label{tmesons}Mass spectra for heavy  mesons;
$q=u,\,d$. 
Units are GeV.}
\begin{center}
%\resizebox{1.05\textwidth}{!}
%{%
\scriptsize{
\begin{tabular}{|c|c||c|c|c|c|c|c|}
\hline
Flavor & Level &  \multicolumn{3}{|c|}{$J=0$} & \multicolumn{3}{|c|}{$J=1$} \\
\cline{3-8}
 & & Particle & Th. mass& Exp. mass  \cite{PDG} & Particle & Th. mass & Exp. mass \cite{PDG} \\
 \hline
$c\bar q$ & $1S$ &  $D$ & $1.862$ &$1.867 $& $D^*$ &2.027
 &$2.008$  \\
& $2S$  & &3.393 &  &   &2.598  &$2.622$  \\
& $3S$  & &   2.837 & &   &  2.987&  \\
\hline
$c\bar s$ & $1S$  & $D_s$ & 1.973&$1.968$ &  $D_s^*$  &2.111 & $2.112$ \\
& $2S$  &&  2.524&  & &2.670 &  \\
& $3S$  &  &2.958&  & &3.064&  \\
\hline $c\bar c$  & $1S$  & $\eta_c$  &  2.990&$2.980$  & $J/\psi$
& 3.125&$3.097$  \\
& $2S$  & &  3.591 &$3.637$ &  &  3.655& $3.686$  \\
& $3S$  &   &3.994 & & &   4.047& $4.039$  \\
\hline
$b\bar q$    & $1S$  & $B$   &5.198 & $5.279$  &   $B^*$     & 5.288&$5.325$  \\
& $2S$  && 5.757&  &  & 5.819&  \\
& $3S$  & &  6.176&  &&  6.220&  \\
\hline
$b\bar s$    & $1S$   & $B_s$ &5.301& $5.366$  & $B_s^*$&5.364&$5.412$  \\
& $2S$  & &  5.856 &  &  &  5.896&  \\
& $3S$  && 6.266&  & & 6.296&  \\
\hline
$b\bar c$ & $1S$ & $B_c$ &6.310& $6.286 $&$B_c^*$&  6.338 &$6.420$ \\
& $2S$  & & 6.869&  &  &   6.879 & \\
& $3S$  & & 7.221 &  &   &      7.228 &  \\
\hline $b\bar b$& $1S$& $\eta_b$&9.387&$9.389$&$
\Upsilon$ & 9.405&$9.460$  \\
& $2S$  && 10.036& && 10.040&$10.023$ \\
& $3S$  &   & 10.369& &  &   10.371&$10.355$\\
& $4S$  &  & 10.619   & & &    10.620&$10.579$ \\
\hline
\end{tabular}
}
\end{center}
\end{table}

The agreement is remarkable, as, for example, in the case of $\eta_b$; in fact, a recent experimental  result is \cite{:2008vj}:
$$M_{\eta_b}=9389^{+3.1}_{-2.3}\pm2.7 \mbox{ MeV}\,.$$

This application can be extended to compute tetraquark masses, in order to examine the possibility that some recently observed puzzling  states, for example, X(3872) and  Y(3940), can be interpreted as diquark-antidiquark bound states. This possibility is justified by the fact that two quarks can attract one another in the one-gluon-exchange approximation, in the $\bar{3}$ color attractive channel.
In the same one-gluon-exchange approximation, the energy of this interaction is half of the energy of the interaction between a quark and an antiquark. Following these hints, we have determined the diquark masses, solving the Salpeter equation (\ref{salpeq}) and using, for the potential $V(r)$,  1/2 $V(r)$ in (\ref{fullpot}). The results  are shown in Table \ref{tdiquark}, in which the square and the curly brackets indicate a diquark with spin 0 and 1, respectively. 
\begin{table}[ht!]
\caption{\small Diquark masses\label{tdiquark}. $\{\}$ denotes a  diquark with spin 1 and $[\, ]$ a diquark with spin 0.
%The model in
%\cite{Ebert:2005nc,Ebert:2007rn} uses a quasipotential of the
%Schr\"odinger type \cite{Ebert:2002pp}.$\{QQ\}$ (resp. $[QQ]$) means
%a spin 1 (resp. $S=0$) diquark $QQ$.  
Units are GeV.}
\begin{center}
\scriptsize{
\begin{tabular}{|c|c||c|c|}\hline
% after \\: \hline or \cline{col1-col2} \cline{col3-col4} ...
 State& Mass & State&
  Mass \\
\hline \{qs\}& 0.980&[qs]&0.979\\
\hline \{ss\}& 1.096&&\\
\hline \{cq\}& 2.168&[cq]&2.120\\
\hline \{cs\}& 2.276&[cs]& 2.235\\
\hline \{cc\}& 3.414&& \\
\hline \{bq\}& 5.526&[bq]&5.513\\
\hline \{bs\}& 5.630&[bs]&5.619\\
\hline \{bc\}& 6.741&[bc]& 6.735\\
\hline \{bb\}& 10.018&& \\
\hline
\end{tabular}
}
\end{center}
\end{table}
 
\begin{figure}[ht]
\begin{center}
\includegraphics[width=6cm]{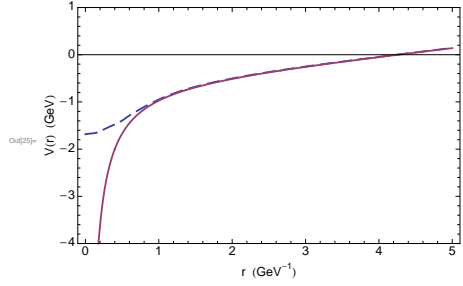}~~~~~~~~~~~~~~~~~~~~~~~~~~~
\end{center}
\caption{\label{modpot} The static energy of a diquark-antidiquark pair
(dashed line) and a quark-antiquark pair (solid
line).  In the first case, the potential corresponds to the
$[cq]\{\bar c\bar q\}$ configuration.}
\end{figure}

We have computed tetraquark masses again solving Eq. (\ref{salpeq}), with now $m_1$ and $m_2$  the masses of the diquark and the antidiquark constituting the tetraquark. The potential $V(r)$ is the same as the one used for mesons, since the interaction is again between states in the representations $\bar{3}$ and $3$ of the SU(3) color group. However, it is necessary to introduce a slight modification, namely an interpolation of the $Q\bar{Q}$ potential (\ref{fullpot}) with the wavefunctions of the diquark  $(\psi_{12}(r))$ and the antidiquark $(\psi_{34}(r))$, due to the fact that diquarks are not point-like particles:
\begin{equation}\label{ddpot}
\tilde V(R)=\frac1N\int d{\bf r_1}\int d{\bf r_2}\,\, |\psi_{12}({\bf r_1})|^2|\psi_{34}({\bf r_2})|^2
V(|{\bf R}+{\bf r_1}-{\bf r_2}|)
\end{equation}
with
\begin{equation}
N=\int d{\bf r_1}\int d{\bf r_2}\, \, |\psi_{12}({\bf r_1})|^2|\psi_{34}({\bf r_2})|^2\,.
\end{equation}
The modified potential $\tilde{V}(r)$ is shown in Fig. \ref{modpot} in the case of the tetraquark $[cq]\{\bar{c}\bar{q}\}$ (dashed line), together with the AdS/QCD potential (\ref{potadsqcd}) (solid line). 

In Table \ref{tethidch}  the masses computed for the tetraquarks with hidden charm are presented. Let us focus on the $1^{++}$ state: according to the hypothesis in \cite{Maiani:2004vq}, it should correspond to the particle X(3872), recently observed, for which the identification with a charmonium state is debated \cite{Choi:2003ue,Jaffe:2004ph}. We find $M=3.899$ GeV, a value not far from experiments. However, confirmation of this interpretation requires the experimental observation of all the other predicted states, for which we obtain the masses reported in Table \ref{tethidch}.

\begin{table}[ht!]
\caption{\small \label{tethidch}Four-quark states with hidden charm.  Units are GeV.
\label{diq}}\begin{center}
{\scriptsize
\begin{tabular}{|c|c|c|c|c|c|}\hline
% after \\: \hline or \cline{col1-col2} \cline{col3-col4} ...
 $J^{PC}$&Flavor content& Mass (this paper)&Mass \cite{Maiani:2004vq}&
  Exp. state &Exp. mass \\
 \hline $0^{++}$&$[cq][\bar c\bar q]$&3.857& 3.723&& \\
 \hline $1^{++}$&$([cq]\{\bar c\bar q\}+[\bar c\bar q]\{cq\})/\sqrt2$
 &3.899& $3.872$& $X(3872)$&$3.8712\pm0.0004$
 \cite{Choi:2003ue}\\
 \hline $1^{+-}$&$([cq]\{\bar c\bar q\}-[\bar c\bar q]\{cq\})/\sqrt2$
 &3.899& $3.754$& &\\
 \hline $0^{++}$&$\{cq\}\{\bar c\bar q\}$&3.729& 3.832& &\\
 \hline $1^{+-}$&$\{cq\}\{\bar c\bar q\}$&3.833& 3.882& &\\
 \hline $2^{++}$&$\{cq\}\{\bar c\bar q\}$&3.988& 3.952& $
 Y(3940)$&$
 3.943\pm0.011\pm0.013$ \cite{Abe:2004zs}\\
\hline
\end{tabular}
}
\end{center}
\end{table}

\section*{Acknowledgements}
I  thank P.~Colangelo, F.~De Fazio and S.~Nicotri  for suggestions. This work was supported in part by the EU Contract No. MRTN-CT-2006-035482, "FLAVIAnet".


\begin{thebibliography}{99}
\bibitem{Karch:2006pv}
  %A.~Karch, E.~Katz, D.~T.~Son and M.~A.~Stephanov,
  A.~Karch {\it et al.},
  \emph{Linear confinement and AdS/QCD},
  \emph{Phys.\ Rev.\  D} {\bf 74}  (2006) 015005;
 % [{ arXiv:hep-ph/0602229}];
  %%CITATION = PHRVA,D74,015005;%%
  %\cite{Colangelo:2007pt}
%\bibitem{Colangelo:2007pt}
  %P.~Colangelo, F.~De Fazio, F.~Jugeau and S.~Nicotri,
  P.~Colangelo {\it et al.},
  \emph{On the light glueball spectrum in a holographic description of QCD},
  \emph{Phys.\ Lett.\  B} {\bf 652} (2007) 73;
%  [{arXiv:hep-ph/0703316}];
  %%CITATION = PHLTA,B652,73;%%
%\cite{Colangelo:2008us}
%\bibitem{Colangelo:2008us}
  %P.~Colangelo, F.~De Fazio, F.~Giannuzzi, F.~Jugeau and S.~Nicotri,
P.~Colangelo {\it et al.},
  \emph{Light scalar mesons in the soft-wall model of AdS/QCD},
  \emph{Phys.\ Rev.\ D} {\bf 78}  (2008) 055009;
%  { arXiv:0807.1054 [hep-ph]};
  %%CITATION = ARXIV:0807.1054;%%
  %\cite{Brodsky:2008pg}
%\bibitem{Brodsky:2008pg}
  S.~J.~Brodsky and G.~F.~de Teramond,
  \emph{AdS/CFT and Light-Front QCD},
  { arXiv:0802.0514 [hep-ph]} and references therein.
  %%CITATION = ARXIV:0802.0514;%%


\bibitem{Andreev:2006ct}
  O.~Andreev and V.~I.~Zakharov,
  \emph{Heavy-quark potentials and AdS/QCD},
  \emph{Phys.\ Rev.\  D} {\bf 74} (2006) 025023.
%  [{ arXiv:hep-ph/0604204}].
  %%CITATION = PHRVA,D74,025023;%%
  
%  \bibitem{Maldacena:1997re}
%  J.~M.~Maldacena,
%  \emph{The large N limit of superconformal field theories and supergravity},
%  \emph{Adv.\ Theor.\ Math.\ Phys.}\  {\bf 2} (1998) 231
%  [Int.\ J.\ Theor.\ Phys.\  {\bf 38} (1999) 1113]
%  [{\tt arXiv:hep-th/9711200}].
%  %%CITATION = IJTPB,38,1113;%%

%

  \bibitem{Carlucci:2007um}
  %M.~V.~Carlucci, F.~Giannuzzi, G.~Nardulli, M.~Pellicoro and S.~Stramaglia,
  M.~V.~Carlucci {\it et al.},
  \emph{AdS-QCD quark-antiquark potential, meson spectrum and tetraquarks},
  { arXiv:0711.2014 [hep-ph]}, to be published in \emph{Eur. Phys. J. C};
  %%CITATION = ARXIV:0711.2014;%%
  F.~Giannuzzi,
  \emph{$\eta_b$ and $\eta_c$ radiative decays in the Salpeter model with the AdS/QCD
  inspired potential},
  arXiv:0810.2736 [hep-ph], to be published in \emph{Phys. Rev. D}.


\bibitem{Maiani:2004vq}
  %L.~Maiani, F.~Piccinini, A.~D.~Polosa and V.~Riquer,
 L.~Maiani  {\it et al.},
 \emph{Diquark-antidiquarks with hidden or open charm and the nature of
  X(3872)},
  \emph{Phys.\ Rev.\  D} {\bf 71} (2005) 014028.
%  [{ arXiv:hep-ph/0412098}].
  %%CITATION = PHRVA,D71,014028;%%
  

\bibitem{Colangelo:1990rv}
  %P.~Colangelo, G.~Nardulli and M.~Pietroni,
P.~Colangelo  {\it et al.},
  \emph{Relativistic Bound State Effects In Heavy Meson Physics},
  \emph{Phys.\ Rev.\  D} {\bf 43} (1991) 3002.
  %%CITATION = PHRVA,D43,3002;%%

\bibitem{PDG}
C.~Amsler {\it et al.} [Particle Data Group], \emph{Phys. Lett. B} {\bf 667} (2008) 1.

\bibitem{:2008vj}
  B.~Aubert {\it et al.}  [BABAR Collaboration],
  \emph{Observation of the bottomonium ground state in the decay $\Upsilon(3S) \rightarrow
  \gamma \eta_b$},
  \emph{Phys.\ Rev.\ Lett.\ }  {\bf 101}  (2008) 071801.
%  [{arXiv:0807.1086 [hep-ex]}].
  %%CITATION = PRLTA,101,071801;%%

\bibitem{Choi:2003ue}
For recent reviews see:
 %P.~Colangelo, F.~De Fazio, R.~Ferrandes and S.~Nicotri,
 P.~Colangelo {\it et al.},
  \emph{Puzzles in charm spectroscopy},
  \emph{Prog.\ Theor.\ Phys.\ Suppl.}\  {\bf 168} (2007) 202;
%  [arXiv:0706.0053 [hep-ph]].
 E.~S.~Swanson,
  \emph{The new heavy mesons: A status report},
  \emph{Phys.\ Rept.}\  {\bf 429} (2006) 243;
%  [arXiv:hep-ph/0601110].
G.~V.~Pakhlova,
  \emph{Exotic ccbar spectroscopy},
  arXiv:0810.4114 [hep-ex].




\bibitem{Jaffe:2004ph}
  R.~L.~Jaffe,
  \emph{Exotica},
  \emph{Phys.\ Rept.}\  {\bf 409} (2005) 1
  [\emph{Nucl.\ Phys.\ Proc.\ Suppl.}\  {\bf 142} (2005) 343].
%  [{arXiv:hep-ph/0409065}].
  %%CITATION = NUPHZ,142,343;%%
  
\bibitem{Abe:2004zs}
  K.~Abe {\it et al.}  [Belle Collaboration],
  \emph{Observation of a near-threshold $\omega \, J/\psi$ mass enhancement in  exclusive
  $B \rightarrow K \omega J/\psi$ decays},
  \emph{Phys.\ Rev.\ Lett.}\  {\bf 94} (2005) 182002.
%  [{arXiv:hep-ex/0408126}].
  %%CITATION = PRLTA,94,182002;%%



\end{thebibliography}
\end{document}